\documentclass[12pt]{article}
\usepackage{amssymb}
\usepackage{amsmath}
\usepackage{amstext}
\usepackage{graphicx,epsfig}
\usepackage{epsfig}
\usepackage{verbatim} 
\usepackage{fancyhdr}
\usepackage{fancybox}
\usepackage{color}
\usepackage{ulem,bbold}
\usepackage{enumitem}
\usepackage{subfigure}
\usepackage{bbm}
\usepackage{parskip}
\usepackage{cite}

\newcommand{\Comment}[1]{{}}
\definecolor{MyDarkBlue}{rgb}{0.15,0.15,0.45}
\usepackage[linktocpage=true]{hyperref}
\hypersetup{
colorlinks=true,
citecolor=MyDarkBlue,
linkcolor=MyDarkBlue,
urlcolor=MyDarkBlue,
}

\newcommand{\be}{\begin{equation}}
\newcommand{\ee}{\end{equation}}
\newcommand{\bea}{\begin{eqnarray}}
\newcommand{\eea}{\end{eqnarray}}
\newcommand{\beas}{\begin{eqnarray*}}
\newcommand{\eeas}{\end{eqnarray*}}
\newcommand{\nn}{\nonumber}

\newcommand{\half}{\frac{1}{2}}

\numberwithin{equation}{section}

\begin{document}


\begin{center}
{\Large \bf{Partially Massless Monopoles and Charges}}
\end{center} 
 \vspace{1truecm}
\thispagestyle{empty} \centerline{
{\large {Kurt Hinterbichler}}$^{a,}$\footnote{E-mail: \Comment{\href{mailto:khinterbichler@perimeterinstitute.ca}}{\tt khinterbichler@perimeterinstitute.ca}} 
{\large  {and Rachel A. Rosen${}^{b,}$}}\footnote{E-mail: \Comment{\href{mailto:rar2172@columbia.edu}}{\tt rar2172@columbia.edu}}
}

\vspace{1cm}

\centerline{{\it ${}^{a}$
Perimeter Institute for Theoretical Physics,}}
\centerline{{\it 31 Caroline St. N, Waterloo, Ontario, Canada, N2L 2Y5 }}
 
 \vspace{1cm}

\centerline{{\it ${}^b$ 
Physics Department and Institute for Strings, Cosmology, and Astroparticle Physics,}}
 \centerline{{\it Columbia University, New York, NY 10027, USA}}

\begin{abstract} 

Massive higher spin fields on de Sitter space exhibit enhanced gauge symmetries at special values of the mass.  These fields are known as ``partially massless."  We study the structure of the charges and Gauss laws which characterize sources for the partially massless spin-2.  Despite having a simple scalar gauge symmetry, there is a rich structure of gauge charges. The charges come in electric and magnetic varieties, each taking values in the fundamental representation of the de Sitter group.  We find two invariant electric-like charges and two invariant magnetic-like charges and we find the point-like monopole solutions which carry these charges,  
analogous to the electric point charge solution and Dirac monopole solution of Maxwell electrodynamics.  These solutions are related by partially massless duality, analogous to the electromagnetic duality that relates electric to magnetic charges.  

\end{abstract}

\newpage

\thispagestyle{empty}
\tableofcontents
\newpage
\setcounter{page}{1}
\setcounter{footnote}{0}

\section{Introduction}
\parskip=5pt
\normalsize

On de Sitter space, massive higher spin fields display the curious phenomenon of partial masslessness \cite{Deser:1983tm,Deser:1983mm,Higuchi:1986py,Brink:2000ag,Deser:2001pe,Deser:2001us,Deser:2001wx,Deser:2001xr,Zinoviev:2001dt,Skvortsov:2006at,Skvortsov:2009zu}.  At particular values of the mass relative to the de Sitter curvature, enhanced gauge symmetries appear.  These gauge symmetries eliminate various lower helicity modes, leaving a field that propagates fewer degrees of freedom than a generic massive field.

The simplest non trivial example is the massive spin-2 field $h_{\mu\nu}$ on de Sitter space,
 \bea \nn S&=&\int d^4x\ \sqrt{-g}\left[ -{1\over 2}\nabla_\lambda h_{\mu\nu} \nabla^\lambda h^{\mu\nu}+\nabla_\lambda h_{\mu\nu} \nabla^\nu h^{\mu\lambda}-\nabla_\mu h\nabla_\nu h^{\mu\nu}+\half \nabla_\mu h\nabla^\mu h\right. \\ &&\left. +3H^2\left( h^{\mu\nu}h_{\mu\nu}-\half h^2\right)-\frac{1}{2}m^2(h_{\mu\nu}h^{\mu\nu}-h^2)\right]. \label{curvedmassivelin}\eea
When the mass takes the following particular value relative to the de Sitter radius $1/H$,
\be m^2={2H^2}\, ,\label{masstuning}\ee
the theory acquires an enhanced scalar gauge symmetry 
\be \delta h_{\mu\nu}=\nabla_\mu\nabla_\nu\, \alpha+H^2 g_{\mu\nu}\,\alpha\, ,\label{gaugesym}\ee
where $\alpha(x)$ is the scalar gauge parameter.
A generic massive spin-2 propagates five physical degrees of freedom, which in the massless limit decomposes into two helicity-2 components, two helicity-1 components and one helicity-0 component.  A partially massless spin-2, with the mass given by \eqref{masstuning}, propagates only 4 degrees of freedom, because the gauge symmetry \eqref{gaugesym} removes the helicity-0 component.

The partially massless spin-2 theory possesses many properties reminiscent of ordinary Maxwell electrodynamics.  
Free Maxwell theory is invariant under a duality symmetry, which acts to interchange the field strength with its Hodge dual: $\delta F_{\mu\nu}=\tilde F_{\mu\nu}$.  Maxwell theory admits point sources with electric charge, which can be detected by Gaussian surfaces which are integrals of the dual field strength over 2-surfaces.   It also admits point sources with magnetic charge, the famous Dirac monopoles, which can be detected by Gaussian surfaces which are integrals of the field strength over 2-surfaces.  Whereas the electric point charge solution can be represented globally in terms of a gauge potential, the magnetic monopole cannot.  The best one can do is to use two potentials in two separate patches which together cover the space, and which in their overlap are related by a gauge transformation.  
In the Maxwell theory the structure of the gauge charges is simple; a single Lorentz invariant number characterizes the electric charge, and another the magnetic charge (with a quantum mechanical quantization condition restricting their product to be proportional to an integer). 

Here, we study the analogous story for the partially massless spin-2 theory.   In the partially massless theory, we will see that the structure of the gauge charges is richer.  The electric charge and magnetic charge are each characterized by a five component vector taking values in the fundamental representation of the de Sitter isometry group $SO(4,1)$.  The invariant notion of charge is the de Sitter invariant norm of this vector, and there are two distinct possibilities corresponding to the cases where the charge vector is spacelike or timelike.

We next find point-like solutions of the partially massless equations of motion which carry these charges.  There are electrically charged solutions, which are the analog of the Coulomb point charge in Maxwell.  We find a two parameter family of such solutions; one parameter carries charge vectors which are spacelike and the other parameter carries charge vectors which are timelike.  Then there are magnetically changed solutions, which are the analog of the Dirac monopole in Maxwell.  Again we find a two parameter family of these solutions.  
As is the case in Maxwell electromagnetism, the electric point charge solution can be represented globally in terms of a partially massless gauge potential $h_{\mu\nu}$, but the magnetic solution cannot; the best one can do is to use two potentials in two separate patches which together cover the space, related in their overlap by a partially massless gauge transformation \eqref{gaugesym}.

\section{Partially Massless Symmetries and Charges}

We consider the partially massless spin-2 theory governed by \eqref{curvedmassivelin} with the mass value \eqref{masstuning}.
Like the photon, the partially massless spin-2 theory possesses a one derivative field strength tensor \cite{Deser:2006zx}
\be F_{\mu\nu\lambda}=\nabla_\mu h_{\nu\lambda}- \nabla_\nu h_{\mu\lambda}.\label{fieldstrengthh}\ee
It is anti-symmetric in the first two indices, and vanishes upon anti-symmetrization of all three indices.  It is invariant under the partially massless gauge symmetry \eqref{gaugesym}.

The action \eqref{curvedmassivelin} can be compactly written in terms of this tensor,
\be\label{Flagr}
S =-\tfrac{1}{4} \int d^4x\ \sqrt{-g}\  \left(F^{\lambda\mu\nu}F_{\lambda\mu\nu}-2F^{\lambda\mu}_{\ \ \ \mu} F_{\lambda\nu}^{\ \ \nu}\right)\, ,
\ee
and the equations of motion away from sources can be cast as the system
\be \nabla_\nu F^{\mu\nu\lambda}=0,\ \ \ F^{\lambda\mu}_{\ \ \ \mu}=0\, .\label{PMFfieldeqs}
\ee
In other words, the on-shell field strength tensor is divergenceless in its anti-symmetric indices, and fully traceless.

\subsection{Partially Massless Duality}

Partially massless fields exhibit a duality invariance analogous to electric-magnetic duality.  This duality was shown to be a symmetry of the action in \cite{Deser:2013xb}, and is displayed in its manifestly covariant form in \cite{Hinterbichler:2014xga}\footnote{Duality exists in other free massless fields, including higher $p$-forms \cite{Teitelboim:1985ya,Teitelboim:1985yc,Nepomechie:1984wu}, linearized gravity \cite{Deser:1981fr,Henneaux:2004jw,Leigh:2007wf} and higher spins \cite{Hull:2001iu,Bekaert:2002dt,Boulanger:2003vs,Deser:2004xt,Deser:2014ssa}.}.
The partially massless duality symmetry acts to interchange the field strength \eqref{fieldstrengthh} with the dual field strength defined by dualizing over the antisymmetric indices\footnote{Here and below, $\epsilon_{\mu\nu\alpha\beta}$ is the standard curved space epsilon {\it tensor} for the spacetime background we are considering,  $\epsilon_{\mu\nu\alpha\beta}=\sqrt{-g}\tilde\epsilon_{\mu\nu\alpha\beta}$ with $\tilde\epsilon_{\mu\nu\alpha\beta}$ the totally antisymmetric epsilon {\it symbol} with $\tilde\epsilon_{0123}=1$.}
\be
\tilde{F}_{\mu\nu}^{~~\lambda} \equiv \frac{1}{2} \epsilon_{\mu\nu\alpha\beta} F^{\alpha\beta\lambda} \, . \label{dualfieldse}
\ee

The partially massless field strength \eqref{fieldstrengthh} satisfies a differential Bianchi identity $\nabla_{[\rho}F_{\mu\nu]\lambda}=0.$  In addition, it satisfies the algebraic identity $F_{[\mu\nu\lambda]}=0$.
 By contracting with the epsilon tensor these two identities can be cast in terms of the dual field strength, where they become respectively
\be \nabla_\nu \tilde F^{\mu\nu\lambda}=0,\ \ \ \tilde F^{\lambda\mu}_{\ \ \ \mu}=0\, .
\ee

These Bianchi identities along with the field equations \eqref{PMFfieldeqs} form a set manifestly invariant under the duality transformation
\be \delta F_{\mu\nu\lambda}=\tilde F_{\mu\nu\lambda}.\ee
As with electromagnetism, duality acts to interchange the field equations with the Bianchi identities.

Duality implies that the theory can be equivalently cast in terms of a dual ``magnetic'' partially massless potential $\tilde h_{\mu\nu}$, non-locally related to the original ``electric'' variable $h_{\mu\nu}$, with its own dual magnetic partially massless gauge symmetry $\delta \tilde h_{\mu\nu}=\nabla_\mu\nabla_\nu\, \tilde\alpha+H^2 g_{\mu\nu}\,\tilde\alpha$, whose field strength is the dual field strength \eqref{dualfieldse}.

\subsection{Partially Massless Charges\label{PMcharges}}

Before looking for point-charge solutions to the partially massless equations of motion, we first identify the conserved charges of the theory which the solutions will carry.  The result is surprisingly rich, in part because the partially massless charges stem from the gauge symmetry \eqref{gaugesym} rather than an ordinary global symmetry.  The symmetries are an example of generalized global symmetries \cite{Barnich:2001jy,Gaiotto:2014kfa}, and lead to 2-form conserved currents.

Non trivial ordinary global symmetries are in one-to-one correspondence via Noether's theorem with non trivial conserved currents, which are 1-forms.  The conserved charge is obtained by integrating the dual of the 1-form current over a $D-1$ dimensional surface.  The conservation of the current and Stokes theorem imply that the value of the integrated charge is independent of deformations of the surface.  The canonical example is to take the surface to be the space-like $t=0$ surface.  Conservation of charge is then the statement that we may deform this surface to any value of $t$ and the result will be the same. 

Gauge symmetries, on the other hand, can lead to 2-form symmetries, which are non trivial conserved 2-form currents.  The associated charges are obtained by integrating the dual of the 2-form current over a $D-2$ dimensional surface, i.e. a Gaussian surface.  Via a generalization of Noether's theorem, non trivial 2-form currents are in one-to-one correspondence with non trivial ``reducibility parameters" of the gauge symmetry.  A reducibility parameter is a value of the gauge parameter for which the gauge transformation vanishes.  The electric and magnetic charges in free Maxwell electromagnetism are the simplest example of this phenomenon\footnote{In theories with only asymptotic reducibilities, such as Yang-Mills and gravity on symmetric backgrounds, the corresponding charges are defined by surface integrals at infinity \cite{Abbott:1981ff,Abbott:1982jh,Barnich:2003xg,Barnich:2007bf}.} (see Appendix \ref{EMcharges}).

To find the reducibility parameters for the partially massless charges, we must find the space of functions $\alpha(x)$ for which the gauge transformation \eqref{gaugesym} is trivial, that is, we must find the general solution to the equation 
\be\nabla_\mu\nabla_\nu\, \alpha+H^2 g_{\mu\nu}\,\alpha\,=0 .\label{gaugesymequ}\ee
It turns out to be quite simple to find the general solution by rewriting \eqref{gaugesymequ} in terms of the ambient-space formulation of de Sitter as a hyperboloid embedded in a five dimensional Minkowski space \cite{Joung:2014aba,Barnich:2015tma}.  In terms of the embedding space coordinates $X^A$ (see Appendix \ref{dscoordapp}), the equation \eqref{gaugesymequ} becomes simply
\be \partial_A \partial_B\alpha(X)=0,\ee
along with an auxiliary condition $X^A\partial_A\alpha(X)=\alpha(X)$.  
The general solution is just a linear combination of the embedding coordinates themselves: $\alpha(X)=C_AX^A$, for some constants $C_A$.  Thus there are five reducibility parameters, which we label by a 5-D Lorentz index.  Pulling back to the de Sitter space we have
\be \alpha^A(x)=X^A(x).\ee
For instance, in global de Sitter coordinates the reducibility parameters are the expressions \eqref{globalembedding}, and in static coordinates they are the expressions \eqref{staticembedding}.

To each reducibility parameter there corresponds a 2-form conserved current.  Generalizing the methods of \cite{Barnich:2001jy} to de Sitter space, we find that this current is given in terms of the gauge invariant field strength \eqref{fieldstrengthh} by
\be j_{\mu\nu}^A=F_{\mu\nu}^{\ \ \lambda}\nabla_\lambda \alpha^A.\label{2formexp}\ee
It is straightforward to see that this is conserved on the equations of motion \eqref{PMFfieldeqs}: differentiating, we have two terms,
$\nabla^\nu j_{\mu\nu}^A= \nabla^\nu F_{\mu\nu}^{\ \ \lambda}\, \nabla_\lambda \alpha^A+F_{\mu}^{\ \nu \lambda}\, \nabla_\nu \nabla_\lambda \alpha^A.$ 
The first term vanishes due to the equation of motion $\nabla^\nu F_{\mu\nu\lambda}=0$.  For the second term, we use the reducibility condition \eqref{gaugesymequ} to write it as $\sim F_{\mu}^{\ \nu \lambda}g_{\nu\lambda} \alpha^A$ after which it vanishes due to the equation of motion $F^{\lambda\mu}_{\ \ \ \mu}=0$.

There are five conserved ``electric'' charges, one for each reducibility parameter, obtained by integrating the dual of the current over some 2-surface
\be Q_E^A=\oint_{S_2} \tilde j^A.\label{eleccgeni}\ee
Like electric charge, these charges are independent of deformations of the 2-surface used to integrate, depending only on the charges enclosed.  Unlike electric charge, they are not invariant under spacetime transformations, rather these charges transform as a vector in the fundamental representation of the de Sitter isometry group $SO(4,1)$.  

Similarly, there are five conserved ``magnetic'' charges, one for each reducibility parameter of the dual ``magnetic'' partially massless symmetry, obtained by integrating the current over some 2-surface
\be Q_M^A=\oint_{S_2}  j^A.\label{magche}\ee
Like the electric charges, they transform in the fundamental representation of the de Sitter isometry group $SO(4,1)$ and are independent of deformations of the 2-surface used to integrate.

In analogy to the Dirac quantization condition, we expect there to be a quantum mechanical consistency condition which results in a de Sitter invariant quantization relation between the electric and magnetic partially massless charges,
\be \eta_{AB} Q_E^AQ_M^B \sim {\rm integer},\ee
where $\eta_{AB}$ is the invariant tensor of the fundamental representation of $SO(4,1)$, which can be thought of geometrically as the metric of the five dimensional Minskowski embedding space.  A similar relation exists in other cases where duality is present \cite{Schwinger:1968rq,Zwanziger:1968rs,Schwinger:1969ib,Deser:1997mz,Deser:1997se,Bunster:2006rt,Barnich:2008ts}.

\section{Monopole Solutions}

We now turn to finding solutions which carry these partially massless charges.  We look for solutions which are regular everywhere except along a singular world-line which can be thought of as a source carrying the charges. 

\subsection{Electric Monopoles}

We start by looking for electric monopole-like solutions which carry only the electric type charges \eqref{eleccgeni}.
We begin by working in static coordinates with the metric \eqref{staticmetric}.  If we look for a spherically symmetric and static solution, $h_{\mu\nu}={\rm diag}\left(f_0(r),f_1(r),f_2(r)r^2,f_2(r)r^2\sin^2\theta\right)$, it is straightforward to show from the partially massless equations of motion that there are no solutions that lead to a non-zero $F$ tensor, so there are no solutions of this type other than pure gauge solutions.  This remains true if we allow for an off-diagonal $h_{tr}$ component depending only on $r$.

To find non trivial solutions, we must allow for a specific time dependence.  Given an ansatz which is a linear combination of the expressions
\be
h_{\mu\nu}=e^{\pm Ht}{\rm diag}\left(f_0(r),f_1(r),f_2(r)r^2,f_2(r)r^2\sin^2\theta\right) \, ,
\ee 
we find the space of solutions\footnote{In fact there is another independent solution 
\be h_{\mu\nu}(t,r)\propto \left(q_0 \cosh(Ht)-q_1 \sinh(Ht)\right){\rm diag}\left(\sqrt{1-H^2 r^2},0,0,0\right),\label{othersol}\ee
but there is also a residual gauge transformation that preserves our ansatz, the transformation \eqref{gaugesym} with 
\be\alpha(t,r)\propto t \left(q_0 \sinh(Ht)-q_1 \cosh(Ht)\right) \sqrt{1-H^2 r^2}.\ee  
This residual symmetry generates \eqref{othersol}, and we may fix the residual symmetry by using it to eliminate this solution.}
\bea
h_{\mu\nu}(x) = {1\over 4\pi}(q_0 \cosh(Ht)-q_1 \sinh(Ht))
\left(
\begin{array}{cccc}
\frac{2\sqrt{1-H^2 r^2}}{ r} &0 &0 &0 \\
0 &-\frac{1}{H^2r^3\sqrt{1-H^2 r^2}}&0 &0 \\
0 &0 &\frac{\sqrt{1-H^2 r^2}}{2 H^2 r} &0 \\
0 &0 &0 &\frac{\sqrt{1-H^2 r^2}}{2 H^2 r}\sin^2\theta
\end{array}
\right) \, , \nn\\ \label{solutioncc}
\eea
which for later convenience we have arranged to be spanned by the given linear combinations with two arbitrary constants $q_0$ and $q_1$.  This solution gives a non trivial field strength $F_{\mu\nu\lambda}$ whose components are given by
\bea F_{tr\mu}&=&{1\over 4\pi}\left(\frac{\sqrt{1-H^2 r^2} }{
   r^2}\left(q_0 \cosh (H t)-q_1 \sinh (H t)\right), \right.  \nn\\
 &&  \left. -\frac{1}{H r^3 \sqrt{1-H^2
   r^2}}\left(q_0 \sinh (H t)-q_1 \cosh (H t)\right),0,0\right),\nn \\
F_{t\theta\mu}&=&{1\over 4\pi}\left(0,0,\frac{\sqrt{1-H^2 r^2}}{2 H
   r}\left( q_0 \sinh (H t)-q_1 \cosh (H t)\right),0\right),\nn \\
F_{t\phi\mu}&=&  {1\over 4\pi}\left(0,0,0,\sin^2\theta\frac{\sqrt{1-H^2 r^2}}{2 H
   r}\left( q_0 \sinh (H t)-q_1 \cosh (H t)\right)\right)\, ,\nn\\
F_{r\theta\mu}&=&{1\over 4\pi}\left(0,0,-\frac{1}{2  \sqrt{1-H^2 r^2}}\left(q_0 \cosh (H t)-q_1 \sinh (H t)\right),0\right)\, ,\nn\\
F_{r\phi\mu}&=&{1\over 4\pi}\left(0,0,0,- \sin^2\theta \frac{1}{2  \sqrt{1-H^2 r^2}}\left(q_0 \cosh (H t)-q_1 \sinh (H t)\right)\right) \, ,\nn \\
F_{\theta\phi\mu}&=&0, \label{fieldstreghte}
\eea
with the other components related by antisymmetry of the first two indices.

We want to compute the conserved charges associated with this solution.  We must evaluate the conserved 2-form current \eqref{2formexp} for various choices of the reducibility parameters described in Section \eqref{PMcharges}, which are given in static coordinates by the expressions \eqref{staticembedding}.  

For 
\be
\alpha^0(r,t) = \sqrt{{1\over H^{2}}-r^2}\, \sinh(H t) \, ,
\ee
we find
\bea
j_{\mu\nu}^0 = 
\left(
\begin{array}{cccc}
0 &-\frac{q_0}{4\pi r^2} &0 &0 \\
\frac{q_0}{4\pi r^2}  &0&0 &0 \\
0 &0 &0 &0 \\
0 &0 &0 &0
\end{array}
\right) \, .
\eea
Despite the non trivial $r$ and $t$ dependence of both the field strength tensor and the reducibility parameter, we see that they conspire to give a two-form current which is independent of time with precisely the geometric $\sim 1/r^2$ dependence required for the surface integral of the dual to be independent of surface.

Integrating the dual $\tilde j^0_{\mu\nu}$ over any two sphere surrounding the origin gives the total ``0-charge" of our solution,
\be Q_E^0=\oint_{S_2} \tilde j^0=q_0.\ee

For 
\be\alpha^1(r,t) = \sqrt{{1\over H^{2}}-r^2}\, \cosh(H t)\ee
we have
\bea
j^1_{\mu\nu} = 
\left(
\begin{array}{cccc}
0 &-\frac{q_1}{4\pi r^2} &0 &0 \\
\frac{q_1}{4\pi r^2}  &0&0 &0 \\
0 &0 &0 &0 \\
0 &0 &0 &0
\end{array}
\right) \, ,
\eea
which gives the ``1-charge,"
\be Q_E^1=\oint_{S_2} \tilde j^1=q_1.\ee

\noindent For the remaining charges we have
\bea
\begin{array}{lcl}
\alpha^2 = r \cos \theta \, , & ~~\Rightarrow~~ & j^2 \propto \cos \theta dr\wedge dt \, , \\
\alpha^3 = r \sin \theta \cos \phi \, , & ~~\Rightarrow~~ & j^3 \propto \sin \theta \cos \phi\, dr\wedge dt , \\
\alpha^4 = r \sin \theta \sin \phi \, , & ~~\Rightarrow~~ & j^4 \propto \sin \theta \sin \phi\, dr\wedge dt ,
\end{array}
\eea
and doing the surface integrals we can straightforwardly see that all the remaining charges are zero,
\be
Q_E^2=Q_E^3=Q_E^4=0.
\ee

We can also compute the magnetic charges \eqref{magche}, and we find that they all vanish,
\be Q_M^A=\oint_{S_2}  j^A=0,\ee
so the solutions we have are purely electric.

As mentioned in Section \ref{PMcharges}, the charges $Q^A_E$ take values in an $SO(4,1)$ vector.  We have found a two-parameter family of solutions, reflecting the freedom for the charge vector to be either timelike or spacelike (with respect to the $SO(4,1)$ invariant $\eta_{AB}$ of the fundamental representation).  The solutions proportional to $q_0$ have a timelike $Q^A_E$ and the solutions proportional to $q_1$ have a spacelike $Q^A_E$.  We can thus obtain a solution with any value of the $Q^A_E$ by performing a de Sitter transformation on one of our solutions.

We have displayed the solutions in static coordinates.  Transforming to global coordinates using \eqref{globtostat}, we find that the components of the gauge potential \eqref{solutioncc} generally blow up at the horizon of the static patch.  However this blow up is pure gauge, because the field strengths \eqref{fieldstreghte} are regular at the horizon and extend smoothly over the entire global de Sitter,
\bea F_{T\chi\mu}&=&{1\over 4\pi}\left({q_0 H\over \sin^2\chi \cosh^2\left(HT\right)},-q_0\tanh\left(HT\right){\cos \chi\over \sin^3\chi}+q_1{1\over \sin^3\chi},0,0  \right),\nn \\
F_{T\theta\mu}&=&{1\over 4\pi}\left(0,0,{q_0\over 2}{\tanh \left(HT\right)\over \tan\chi}-{q_1\over 2 \sin\chi},0\right),\nn \\
F_{T\phi\mu}&=&{1\over 4\pi}\left(0,0,0,\left[{q_0\over 2}{\tanh \left(HT\right)\over \tan\chi}-{q_1\over 2 \sin\chi}\right]\sin^2\theta\right),\nn \\
F_{\chi\theta\mu}&=&  {1\over 4\pi}\left(0,0,-{q_0 \over2 H},0\right)\, ,\nn\\
F_{\chi\phi\mu}&=&  {1\over 4\pi}\left(0,0,0,-{q_0 \over2 H}\sin^2\theta\right)\, ,\nn\\
F_{\theta\phi\mu}&=&\left(0,0,0,0\right) . \label{fieldstreghteglobalc}
\eea
The only places where this field strength blows up are the north and south poles, $\chi=0$ and $\chi=\pi$ respectively, reflecting the presence of the partially massless charge and a compensating mirror charge at the opposite pole.  The presence of this additional charge is necessary because global de Sitter has spatial sections which are compact 3-spheres.  This puts a constraint on the allowed charges.  If we have a charged object at the north pole, then a 2-surface surrounding this charged object can also be interpreted as a 2-surface surrounding all the rest of the 3-sphere.  Thus there must be other charged objects present, such that their total charge balances that of the charge at the north pole.  The total charge of all objects on the spatial 3-sphere must be zero.  Figure \ref{dstosphere} shows a global view of the solution.  (See Appendix \ref{dSmonopole} for the analogous case of electromagnetism on de Sitter.)

\begin{figure}[h!]
\begin{center}
\epsfig{file=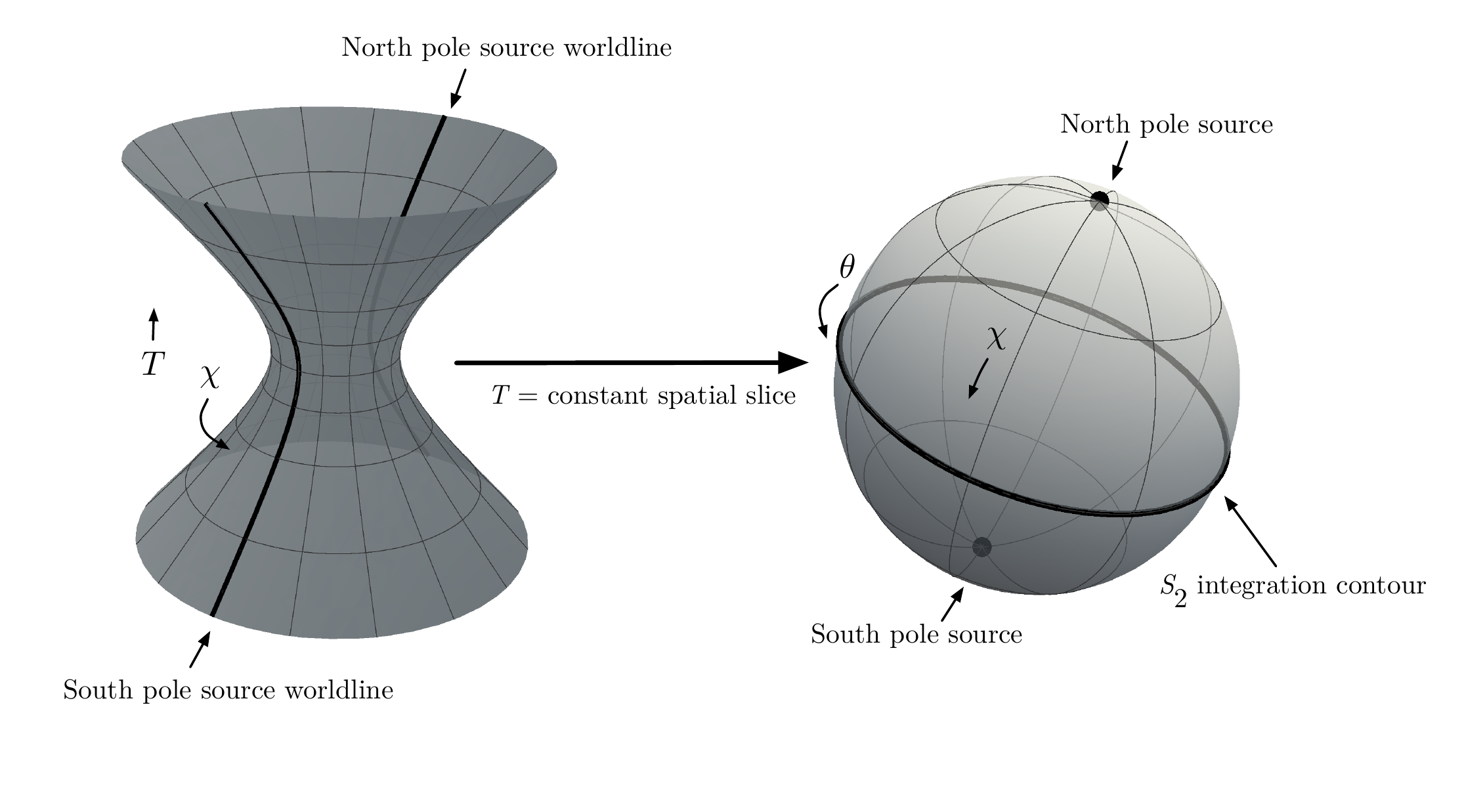,height=3.8in,width=7.0in}
\caption{\small Global view of the monopole on de Sitter.  The left shows the global de Sitter hyperboloid with $T$ going vertically and $\chi$ parametrizing the spherical spatial sections ($\theta$ and $\phi$ suppressed).  The worldlines of the sources lie on the north pole at $\chi=0$ and the south pole at $\chi=\pi$.  The right shows a spatial $S^3$ slice.  $\chi$ is latitude and $\phi$ is longitude ($\phi$ is suppressed).  The $S^2$ integration contour along the equator at $\chi=\pi/2$ can be interpreted as enclosing either of the two charges, so they must be equal and opposite.}
\label{dstosphere}
\end{center}
\end{figure}

In fact, one can find a corresponding potential in global coordinates that is regular everywhere except at the poles, whose field strength yields \eqref{fieldstreghteglobalc},
\bea
h_{TT} &=& \frac{q_0}{4 \pi} \frac{H}{\tan\chi \cosh^2(H T)} \, , \nn\\
h_{\chi\chi} &=& \frac{q_0}{4 \pi} \frac{1+\csc^2 \chi}{H \tan \chi} + \frac{q_1}{2 \pi} \frac{\tan^{-1}(\tanh(H T/2))\cosh(H T)}{H \sin^3 \chi}\, ,\nn \\
h_{\theta\theta} &=& -\frac{q_0}{4 \pi} \frac{1-2 \sin^2 \chi}{2 H \tan \chi} -\frac{q_1}{4\pi}\frac{\tan^{-1}(\tanh(H T/2))\cosh(H T)}{H \sin \chi}\, , \nn\\
h_{\phi\phi} &=& \sin^2\theta\, h_{\theta\theta} \, .
\eea

\subsection{Magnetic Monopoles}

Next we look for magnetic monopole-like solutions which carry only the magnetic type charges \eqref{magche}.
The magnetic solutions should have a field strength which is the dual of the electric solutions' field strength.
Thus we wish to find an $h_{\mu\nu}$ whose field strength is the dual of the $F_{\mu\nu\lambda}$ of the previous section.  As is the case with the Dirac monopole, we will need two different solutions to cover the northern and southern hemispheres.  We find the two parameter family, in static coordinates,
\bea
h_{03}^\pm &=&  {1\over 4\pi} (-g_0 \cosh(Ht)+g_1 \sinh(Ht)){\sqrt{1-H^2 r^2}}(\pm 1-\cos\theta) \, ,\nn \\
h_{13}^\pm &=& {1\over 4\pi} (g_0 \sinh(Ht)-g_1 \cosh(Ht))\frac{1}{Hr\sqrt{1-H^2 r^2}}(\pm 1-\cos\theta)\, ,   \nn \\
h_{23}^\pm &=& {1\over 4\pi}  (-g_0 \sinh(Ht)+g_1 \cosh(Ht)) \frac{\sqrt{1-H^2 r^2}}{2H}\frac{\left(\pm 1-\cos\theta\right)^2}{\sin\theta} \, , \label{maghpote}
\eea
with the other components zero or related by symmetry.  
The solution $h_{\mu\nu}^+$ covers everything except the south poles of the spheres parametrized by $\theta,\phi$, and $h_{\mu\nu}^-$ covers except the north poles.
In the overlap region away from the poles, these solutions are related by a partially massless gauge transformation,
\be
h^+_{\mu\nu}-h^-_{\mu\nu} = \nabla_\mu\nabla_\nu \alpha+H^2 g_{\mu\nu} \alpha \, , ~~~~ \alpha(x) = \, {1\over 2\pi}\left(-g_0 \sinh(Ht)+g_1 \cosh(Ht)\right) \frac{\sqrt{1-H^2 r^2}}{H} \phi \, .
\ee

The field strength of \eqref{maghpote} is the dual of the field strength \eqref{fieldstreghte} with $q_0,q_1\leftrightarrow g_0,g_1$.
The conserved 2-form is simply the dual of the conserved two form of the electric solution,
and it is straightforward to see that the magnetic solutions will have non-vanishing magnetic charges
 and vanishing electric charges
 \bea && Q_M^0=\oint_{S_2}  j^0= g_0,\ \ \ Q_M^1=\oint_{S_2}  j^1 = g_1,\ \ Q_M^2=Q_M^3=Q_M^4=0,
 \eea
\be Q_E^A=\oint_{S_2}  \tilde j^A=0.\ee

\subsection{Sources}
We end this section with a brief discussion of charged sources in the partially massless theory.  We wish to verify that our point charge solutions can indeed be generated by sources in the theory.  Consider the partially massless theory \eqref{Flagr} coupled to a non-dynamical source $T^{\mu\nu}$,
\be
S = \int d^4x\ \sqrt{-g}\ \left[  -\tfrac{1}{4}\left(F^{\lambda\mu\nu}F_{\lambda\mu\nu}-2F^{\lambda\mu}_{\ \ \ \mu} F_{\lambda\nu}^{\ \ \nu}\right)+h_{\mu\nu}T^{\mu\nu}\right]\, ,
\ee
In order for this coupling to maintain the partially massless symmetry \eqref{gaugesym}, the source must obey the following conservation law,
\be
\label{PMcons}
\left(\nabla_\mu\nabla_\nu+H^2 g_{\mu\nu}\right) T^{\mu\nu} = 0 \, .
\ee
To make contact with the previous sections, let us look for solutions to this equation for an isotropic source.  Taking a spherically symmetric ansatz with a time dependence mirroring that of our monopole solutions, we find
\bea
\label{Tmunu}
T^{\mu\nu} = \left(c_0 \cosh Ht-c_1 \sinh Ht \right)
\left(
\begin{array}{cccc}
f(r)+\frac{1}{4}r\,f'(r) &0 &0 &0 \\
0 &\frac{1}{4} H^2 r^2 (1-H^2 r^2) f(r)&0 &0 \\
0 &0 &0 &0 \\
0 &0 &0 &0
\end{array}
\right) \, .
\eea
For a generic function $f(r)$ and parameters $c_0$ and $c_1$, this satisfies the partially massless conservation equation\footnote{So long as $f(r)$ is not proportional to $\frac{1}{r^4 \sqrt{1-H^2 r^2}}$, this satisfies \eqref{PMcons} in a non trivial way, meaning that the trace and double divergence do not separately vanish.} \eqref{PMcons}. 

In the presence of a source, the equations of motion no longer yield $F^{\lambda\mu}_{\ \ \ \mu}=0$ and the conserved current \eqref{2formexp} is modified,
\be
j_{\mu\nu}^{\ A} = F_{\mu\nu}^{\ \ \lambda}\nabla_\lambda \alpha^A-F_{\mu\lambda}^{\ \ \lambda} \nabla_\nu\alpha^A+F_{\nu\lambda}^{\ \ \lambda}\nabla_\mu \alpha^A \, .
\ee
On-shell, the divergence of this 2-form current can be written in terms of the source as
\be
\label{onecurrent}
j_\mu^{\ A} \equiv \nabla^\nu j_{\mu\nu}^{\ A} = T_{\mu\nu}\nabla^\nu \alpha^A-(\nabla^\nu T_{\mu\nu}) \,\alpha^A \, .
\ee
Using the conservation equation \eqref{PMcons}, it is straightforward to see that the divergence of this expression vanishes: $\nabla^\mu j_\mu^{\ A} = 0$.

If we plug our isotropic source \eqref{Tmunu} into the above expression we find that
\bea
j_\mu^{\ 0} = - c_0 \left((f(r)+\tfrac{1}{4}r f'(r))(1-H^2 r^2)^{3/2},0,0,0\right) \, , \\
j_\mu^{\ 1} = - c_1 \left((f(r)+\tfrac{1}{4}r f'(r))(1-H^2 r^2)^{3/2},0,0,0\right) \, .
\eea
Again we see that the currents are time independent, in spite of the non trivial time dependence of the source \eqref{Tmunu}.  In particular, for a point-like source, i.e. for
\be
f(r) \propto \frac{\delta(r)}{r^2} \, ,
\ee
we see that we can identify the constants $c_0$ and $c_1$ with the charges $q_0$ and $q_1$ respectively, up to an overall factor.  This is because the volume integrals of $j_\mu^{\ A}$ are equal to the surface integrals of $j_{\mu\nu}^{\ A}$.

\section{Flat Space Limit}
We can better understand the partially massless charges by considering the flat space limit of the theory.  We take this limit so that both $m$ and $H$ go to zero in a way that preserves the necessary condition for the partially massless symmetry $m^2 = 2H^2$.  The resulting theory is that of a free, massless spin-2 field and and a free, massless spin-1 field.  This theory has an enhanced gauge symmetry compared to the partially massless theory: it has linearized diffeomorphism invariance as well as the usual $U(1)$ gauge symmetry of electromagnetism.

To study what happens to the partially massless charges, we can consider the equation \eqref{gaugesymequ} for the reducibility parameters in this limit,
\be
\nabla_\mu\nabla_\nu\, \alpha+H^2 g_{\mu\nu}\,\alpha\,=0 ~~~~\Rightarrow~~~~ \partial_\mu\partial_\nu \alpha = 0 \, .
\ee
The five solutions to this equation are given by $\alpha^\mu = x^\mu$ and $\alpha = ~const$.  The first four of these are a subset of the gauge charges associated with linearized diffeomorphism invariance.  The equation for the reducibility parameters of linearized diffeomorphism invariance is $\partial_{(\mu}\xi_{\nu)} = 0$,
which has the solutions $\xi_\mu = \partial_\mu \alpha$ with $\alpha^\nu = x^\nu$.  These are the four reducibility parameters associated with 4-momentum conservation.  Thus we can associated four of the partially massless charges with the 4-momentum in the flat space limit.  The invariant sum of the squares of the charges would of course be the mass.

The remaining reducibility parameter $\alpha = ~const$ is not associated with diffeomorphism invariance since it would correspond to $\xi_\mu = 0$.  Instead, it is associated with the gauge charge of the $U(1)$ symmetry, i.e., with the electric charge, as the reducibility equation is $\partial_\mu \alpha =0$ .

By considering the sourced theory, we can readily see that, in the flat space limit, the ``timelike" partially massless charge $q_0$ is associated with the mass while the ``spacelike" charge $q_1$ is associated with the electric charge.  We perform the usual St\"uckelberg trick (see, e.g., \cite{Hinterbichler:2011tt}) in order to isolate the helicity-2 and helicity-1 modes of the partially massless graviton (note that we need not introduce a helicity-0 field because it is eliminated by the partially massless symmetry),
\be
h_{\mu\nu} \rightarrow h_{\mu\nu} +\frac{1}{2H}\nabla_{(\mu}A_{\nu)} \, .
\ee
Here we have canonically normalized the helicity-1 field $A_\mu$.  The coupling to $T^{\mu\nu}$ is thus:
\be
h_{\mu\nu} T^{\mu\nu} \rightarrow h_{\mu\nu} T^{\mu\nu}+\frac{1}{2H}\nabla_{(\mu}A_{\nu)} T^{\mu\nu} \, ,
\ee
which we can write as
\be
h_{\mu\nu} T^{\mu\nu}+A_\mu J^\mu \, , 
\ee
with
\be
\label{Jmu}
J^\mu \equiv - \frac{1}{H}\nabla_{\nu}T^{\mu\nu} \, .
\ee
If we then take the flat space limit of the sources \eqref{Tmunu} and \eqref{Jmu}, we can identify
\be
T^{\mu\nu} \rightarrow c_0
\left(
\begin{array}{cccc}
f(r)+\frac{1}{4}r\,f'(r) &0 &0 &0 \\
0 & 0 &0 &0 \\
0 &0 &0 &0 \\
0 &0 &0 &0
\end{array}
\right) \, ,
\ee
\be
J^\mu \rightarrow c_1 \left(
f(r)+\frac{1}{4}r\,f'(r), 0, 0, 0 \right) \, .
\ee
Thus, at least in the context we are considering here, it appears that we can identify $q_0$ with mass and $q_1$ with electric charge in the flat space limit.

\section{Discussion}

The partially massless spin-2 theory has been of interest as a possible gravitational theory.  If our graviton were described by a partially massless spin-2, the relation \eqref{masstuning} which is enforced by the partially massless gauge symmetry \eqref{gaugesym} would fix the value of the cosmological constant relative to the value of the mass of the graviton, which can itself be small in a technically natural sense due to the enhanced diffeomorphism invariance of general relativity at $m=0$ \cite{deRham:2012ew,deRham:2013qqa}.   However, there are various no-go results which make it difficult to realize a fully non-linear theory containing a partially massless mode \cite{Zinoviev:2006im,Deser:2012qg,deRham:2013wv,Hinterbichler:2013kwa,Joung:2014aba,Zinoviev:2014zka,Garcia-Saenz:2014cwa}.

In the case of Maxwell electromagnetism, a non-linear realization is given by an $SU(2)$ gauge theory spontaneously broken to $U(1)$ by the VEV of an adjoint Higgs.  The Dirac monopole of the $U(1)$ theory is resolved into the 't Hooft-Polyakov monopole of the $SU(2)$ theory \cite{tHooft:1974qc,Polyakov:1974ek}.  It might be expected that a potential non-abelian partially massless theory which realizes or completes a partially massless spin-2 would also contain solitonic solutions which resolve the monopole solutions found here.

\vskip.5cm

\bigskip
{\bf Acknowledgements}: 
The authors would like to thank Alberto Nicolis for helpful discussions. Research at Perimeter Institute is supported by the Government of Canada through Industry Canada and by the Province of Ontario through the Ministry of Economic Development and Innovation. This work was made possible in part through the support of a grant from the John Templeton Foundation. The opinions expressed in this publication are those of the authors and do not necessarily reflect the views of the John Templeton Foundation (KH).  RAR is supported by DOE grant DE-SC0011941.  RAR would like to thank the Perimeter Institute for hospitality while this work was underway.

\appendix
\section{Coordinates on $dS_4$\label{dscoordapp}}

Four dimensional De Sitter space of radius ${1/H}$ can be described as the subset of points embedded in a five dimensional Minkowski space, $(X^0,X^1,X^2,X^3,X^{4})\in\mathbb{R}^{1,4}$, the hyperbola of one sheet satisfying 
\be -(X^0)^2+(X^1)^2+(X^2)^2+(X^3)^2+(X^{4})^2={1\over H^2},\ee
with the metric induced from the usual flat Minkowski metric on $\mathbb{R}^{1,n}$.  The line of points intersecting the planes $X^0,X^2,X^3,X^4=0$ and having $X^1>0$ is called the \textit{north pole}, those having $X^1<0$ the \textit{south pole}.
The scalar curvature $R$ and cosmological constant $\Lambda$ are 
\be R={12  { H}^2},\ \ \ \Lambda={3 { H}^2}.\ee
Global coordinates cover the entire space and are given by the embedding 
\bea \nn X^0&=&{1\over H}\sinh\left(H T\right), \\ \nn
X^1&=&{1\over H}\cosh\left(H T\right)\cos\chi, \\ \nn
X^2&=&{1\over H}\cosh\left(H T\right)\sin\chi\cos\theta, \\ \nn
X^3&=&{1\over H}\cosh\left(H T\right)\sin\chi\sin\theta\cos\phi ,\\
X^4&=&{1\over H}\cosh\left(H T\right)\sin\chi\sin\theta\sin\phi . \label{globalembedding}
\eea
The coordinate ranges are 
$T\in (-\infty,\infty)$, $\chi\in (0,\pi)$, and $(\theta,\phi)\in S_2$.  
The metric is
\be\label{desitterglobal} {ds^2=-dT^2+{1\over H^2}\cosh^2\left(H T\right)\left[d\chi^2+\sin^2\chi\left(d\theta^2+\sin^2\theta d\phi\right)\right].}\ee

$T$ is a time coordinate, and $\chi,\theta,\phi$ are angles on a spatial 3-sphere, where constant $\chi$ lines are $S_2$'s parametrized by $\theta,\phi$.  The north pole is at $\chi=0$ and the south pole is at $\chi=\pi$.

Static coordinates cover only the region $X^1>|X^0|$.  The embedding is given by 
\bea \nn X^0&=&\sqrt{{1\over { H}^2}-r^2}\sinh\left(Ht\right)\\ \nn
X^1&=&\sqrt{{1\over { H}^2}-r^2}\cosh\left(H t \right)\\ \nn
X^2&=&r\cos\theta\\ \nn
X^3&=&r\sin\theta\cos\phi\\ 
X^4&=&r\sin\theta\sin\phi.  \label{staticembedding} 
\eea
The coordinates ranges are $t\in (-\infty,\infty)$, $r\in (0,1/H)$, and $(\theta,\phi)\in S_2$.
The metric is
\be \label{staticmetric}{ ds^2=-\left(1-{H^2 r^2}\right)dt^2+{1\over 1-{H^2r^2}}dr^2+r^2\left(d\theta^2+\sin^2\theta d\phi\right).}\ee
The north pole is at $r=0$, and the horizon of the static patch is at $r=1/H$.

\begin{figure}[h!]
\begin{center}
\epsfig{file=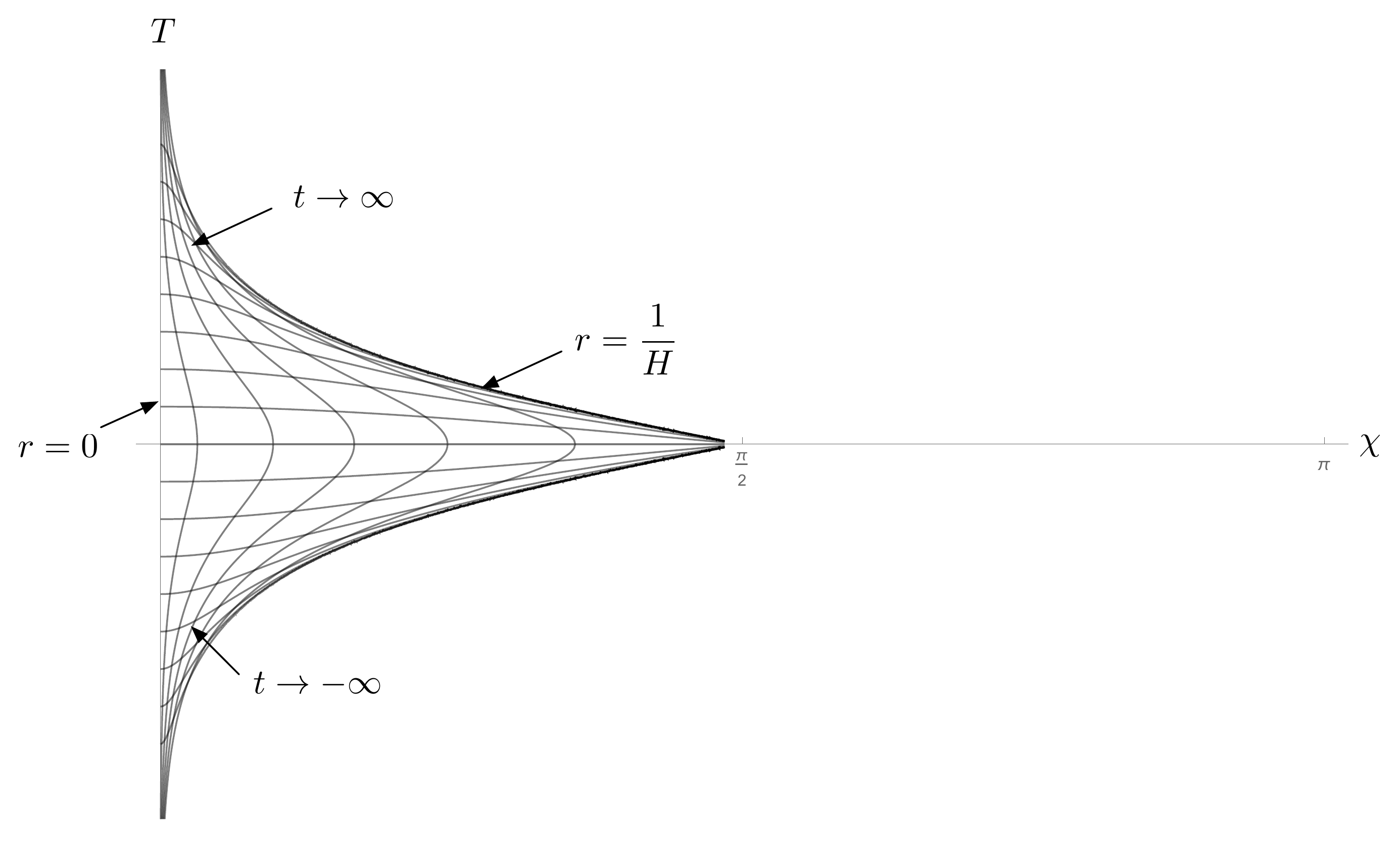,height=3in,width=4.5in}
\caption{\small Static coordinates as seen in global coordinates.}
\label{staticinglobalc}
\end{center}
\end{figure}

The static coordinates are embedded in the global coordinates as
\be t={1\over H}\tanh^{-1}\left(\sec\chi\tanh(HT)\right),\ \ \ r={1\over H}\cosh(HT)\sin\chi,\ \ \ \theta=\theta,\ \ \ \phi=\phi,\label{globtostat}\ee
see Figure \ref{staticinglobalc}.

\section{Review of Electromagnetic Monopoles and Charges}
\subsection{Electromagnetic Charges}\label{EMcharges}
Electric and magnetic charges in free Maxwell electromagnetism are not conserved charges corresponding to an ordinary symmetry via Noether's theorem.  Rather they are 2-form symmetries, as described in Section \ref{PMcharges}.

Maxwell electromagnetism has the gauge symmetry
\be \delta A_\mu=\partial_\mu\Lambda,\label{emgauge}\ee
with scalar gauge parameter $\Lambda(x)$.  The reducibility parameters are the values of $\Lambda(x)$ for which \eqref{emgauge} vanishes.  These are simply the constant values of the gauge parameter, $\Lambda=const.$  Thus the space of reducibility parameters is one dimensional, and there is one associated non trivial conserved 2-form current which happens to be the field strength $F=dA$,
\be j^E= F.\ee
It is conserved by virtue of the source free Maxwell's equations $\partial^\mu F_{\mu\nu}=0$.  The electric charge of a source is then given by 
\be Q_E=\oint_{S_2}\tilde F\ ,\ee
where $S_2$ wraps the world-volume of the source and $\tilde F_{\mu\nu}\equiv {1\over 2}\epsilon_{\mu\nu}^{\ \ \ \rho\sigma}F_{\rho\sigma}$ is the Hodge dual.

In four dimensions, electromagnetism possesses electromagnetic duality \cite{Dirac:1931kp,Dirac:1948um}.  Electromagnetic duality is a symmetry of the sourceless Maxwell theory \cite{Deser:1976iy,Deser:1981fr}.  When acting on the field strength, duality symmetry acts to transform the field strength to its dual, $\delta F_{\mu\nu}=\tilde F_{\mu\nu}$.
There is a magnetic gauge potential $\tilde A$ whose field strength is $\tilde F$, and a dual gauge symmetry which acts on $\tilde A$.  Associated with the reducibility parameter of the dual gauge symmetry, there is a magnetic 2-form current given by the dual of the electric 2-form current,
\be j^M= \tilde F.\ee
It is conserved by virtue of the Bianchi identities of the free Maxwell's equations $dF=0\rightarrow\partial^\mu \tilde F_{\mu\nu}=0$.  The magnetic charge of a source is then given by 
\be Q_M=\oint_{S_2} F\ ,\ee
where $S_2$ wraps the world-volume of the source. 

There are basic point-like solutions of the Maxwell equations which can be thought of as Wilson lines or 't Hooft lines in the theory, charged under the electric or magnetic symmetries respectively.  These solutions are the electric point charge and Dirac monopole solutions, which we review next.

\subsection{Flat Space Monopoles}

For the electric solution in flat Minkowski space, we have, in spherical coordinates,
\be F={q\over 4\pi r^2}dr\wedge dt,\label{elecsol1}\ee
with $q$ a constant.  Using $\tilde F=q{\sin\theta\over{4\pi} }d\theta\wedge d\phi$, the electric charge is
\be Q_E=\oint_{S_2}\tilde F=q\oint_{S_2} {\sin\theta\over{4\pi}} d\theta\wedge d\phi=q,\ee
and the magnetic charge vanishes,
\be Q_M=\oint_{S_2} F=0.\ee
The solution can be represented globally as $F=dA$ using the gauge potential
\be A=-{1\over 4\pi r}dt.\ee

For the magnetic solution, we have, in spherical coordinates,
\be F=g{\sin\theta\over{4\pi} }d\theta\wedge d\phi.\ee
This is the Hodge dual of the electric solution \eqref{elecsol1}, with $q\leftrightarrow g$.  
The magnetic charge is
\be Q_M=\oint_{S_2} F=g.\ee
and the electric charge vanishes,
\be Q_E=\oint_{S_2} \tilde F=0.\ee

As is well known, the magnetic solution cannot be represented globally by a single electric gauge potential (i.e. $F$ is closed but not globally exact).  Instead we need at least two potentials defined in two charts that together cover the space, for example
\be A^\pm={g\over 4\pi }\left(\pm1-\cos\theta\right)d\phi.\ee
$A^+$ covers everything except the south pole of the $\theta,\phi$ spheres, and $A^-$ covers everything except the north pole.
In the overlap region they differ by a gauge transformation,
\be A^+ -A^-=d\Lambda,\ \ \ \ \Lambda={g\over 2\pi}\phi.\ee

Quantum mechanically, there is a Dirac quantization condition that fixes the product of the possible electric and magnetic charges of sources to be integers\footnote{One way to get the famous Dirac quantization condition is to couple in a point particle or a Wilson line, with action $S=e\oint A=e\int d\tau A_\mu(X(\tau)){dX^\mu\over d\tau}$, where the integral is over the world-line $X^\mu(\tau)$.  We impose the condition that the path integrand, $e^{iS}$, be gauge invariant, and hence independent of the choice of $A$.  This requires
\be S[A^+-A^-]=2\pi n,\ \ \ n\in {\rm integers}.\ee
Taking the wilson line to run around the equator of the sphere, i.e. $X^t(\tau)=X^r(\tau)=0$, $X^\theta (\tau)={\pi\over 2}$, $X^\phi(\tau)=\tau$, we have 
\be  S[A^+-A^-]=e\int_0^{2\pi} d\tau \ (d\Lambda)_\phi =e{g\over 2\pi}\left.\tau \right|^{2\pi}_0=eg,\ee
and so the quantization condition is
\be eg=2\pi n\, , \ \ \ n\in {\rm integers}.\ee},
\be eg=2\pi n\, ,\ \ \ n\in {\rm integers}.\ee

\subsection{de Sitter Space Monopoles}\label{dSmonopole}
The partially massless fields live on de Sitter space, so to make a better comparison we now find the de Sitter space versions of the Maxwell electric and magnetic monopoles.  (The corresponding non-abelian versions have been extensively studied \cite{vilenkin2000cosmic,Vachaspati:2000cq}.)  They show new non trivial features due to the closed spatial slices of global de Sitter.

In static coordinates (see Appendix \ref{dscoordapp} for a review and conventions on de Sitter coordinate systems), the electric solution is 
the most general static, spherically symmetric solution centered on the north pole,
\be A=-{q\over 4\pi r}dt,\ \ \  F={q\over 4\pi r^2}dr\wedge dt.\label{elecsole}\ee
The electric and magnetic charges are, respectively
\be Q_E=\oint_{S_2}\tilde F=q,\ \ Q_M=\oint_{S_2} F=0.\ee
The corresponding magnetic solution again requires two patches to cover,
\be A^\pm={g\over 4\pi }\left(\pm 1-\cos\theta\right)d\phi\, ,\ \ \ \ F=g{\sin\theta\over{4\pi} }d\theta\wedge d\phi. \ee
The magentic and electric charges are, respectively
\be Q_M=\oint_{S_2} F=g,  \ \ Q_E=\oint_{S_2} \tilde F=0.\ee

\begin{figure}[h!]
\begin{center}
\epsfig{file=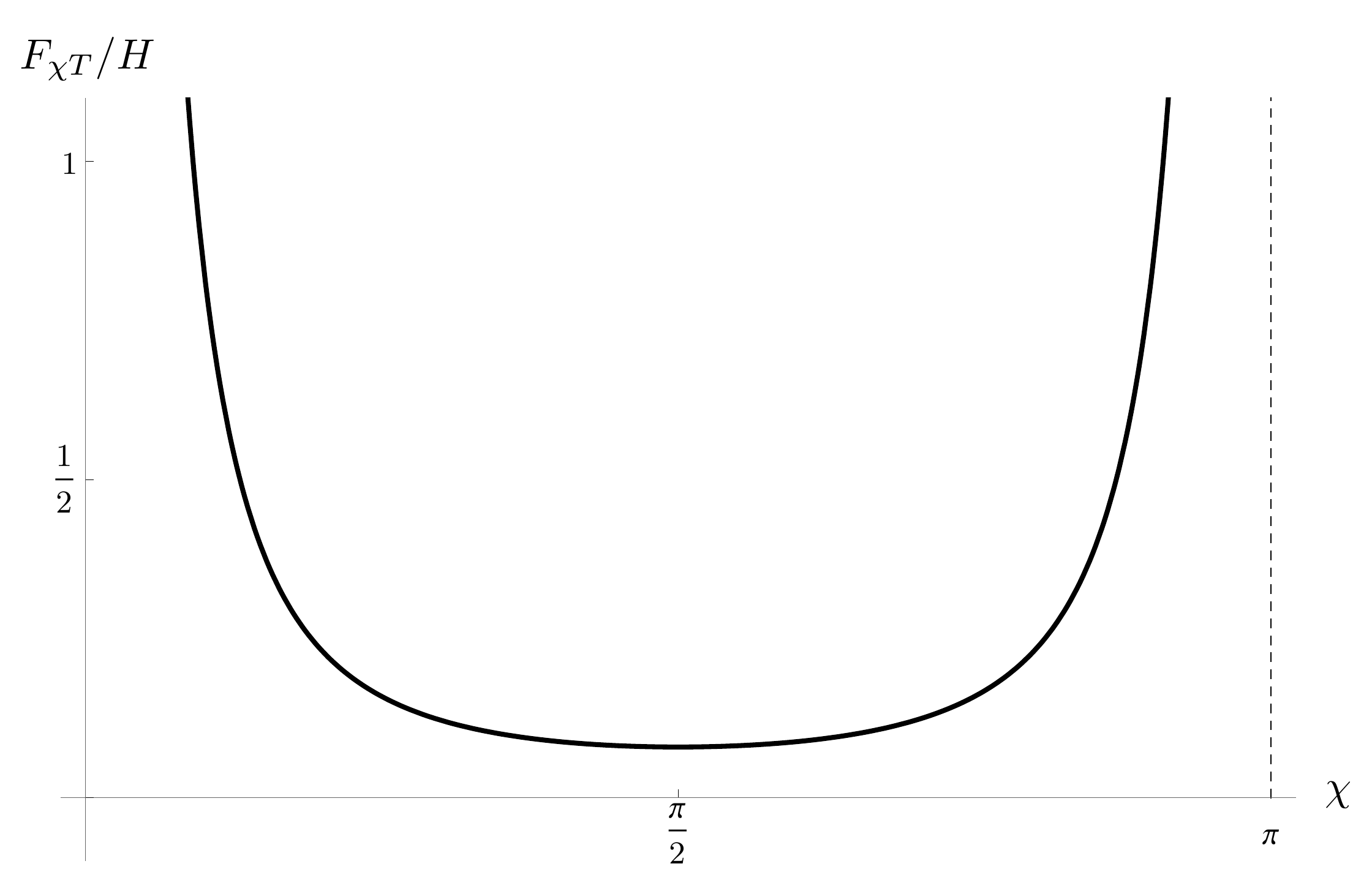,height=3in,width=4.5in}
\caption{\small Electric field component $F_{\chi T}$ in global coords at $T=0$.  We see the blow up at the north and south poles, $\chi=0$ and $\chi=\pi$ respectively, corresponding to the presence of equal and opposite electric sources.}
\label{felecglobsol}
\end{center}
\end{figure}

The field strengths of the static patch point charge solutions above extend to solutions over the global de Sitter.   
The electric solution's potential \eqref{elecsole} changed to global coordinates using \eqref{globtostat} reads
\be A={H\over 4\pi\left({\tanh^2(HT)\over \cos^2\chi}-1\right)}\left[{1\over \cos\chi\sin\chi\cosh^3(HT)}dT+{\tanh(HT)\over \cosh(HT)\cos^2\chi}d\chi\right].\ee
This solution blows up at the horizon of the static patch.  But this blow-up is pure gauge, because the field strength becomes
\be F=-{H\over 4\pi \cosh(HT)\sin^2\chi}dT\wedge d\chi,\ee
which is regular at the horizon and is well defined over the entire global de Sitter space except for the north and south poles.  The original charge is at the north pole, and we see that there is another source at the south pole, with an equal and opposite charge, satisfying the Gauss-law constraint that the total charge on the compact spatial 3-spheres must be zero (see Figures \ref{dstosphere} and \ref{felecglobsol}).

The magnetic solution depends only on $\theta,\phi,d\theta,d\phi$, and so the expression goes unchanged into global dS,
\be A^\pm={g\over 4\pi }\left(\pm 1-\cos\theta\right)d\phi,\ \ \ F=g{\sin\theta\over{4\pi} }d\theta\wedge d\phi. \ee
Again, there is a magnetic charge at the south pole which compensates for the magnetic charge at the north pole.

\bibliographystyle{utphys}
\addcontentsline{toc}{section}{References}
\bibliography{PMmonopole_arxiv2}

\end{document}